\documentstyle[11pt]{article}
\begin{document}

\title{\bf Stochastic Coefficient of Restitution --- a New Approach to the
Dynamics of  Granular Media}

\author{G\"otz Giese\\
Institut f\"ur Theoretische Physik, Universit\"at G\"ottingen,\\
Bunsenstr.~9,
37073 G\"ottingen,
Germany\thanks{e-mail:giese@theorie.physik.uni-goettingen.de}}

\date{\today}
\maketitle

\begin{abstract} We consider a one--dimensional "gas" of
inelastically colliding particles where kinetic energy is dissipated
by the excitation of vibrational degrees of freedom. In our model the
coefficient of restitution is a stochastic quantity whose
distribution can be calculated from an exact stochastic equation of
motion. We investigate the equipartition properties of the system and
propose a new algorithm for computer simulations, that is a
combination of event--driven(ED) and Monte--Carlo methods.
\end{abstract}

Numerical and theoretical approaches to the dynamics of granular materials
frequently adopt the concept of a coefficient of restitution that
determines the energy loss during collisions of granular particles.
Event--driven (ED) simulations\cite{Bernu,McNamara,Luding1} have shown
that model systems with {\em fixed} coefficient of restitution evolve into
clustered states where a hydrodynamic description ceases to be correct:
Fundamental assumptions of hydrodynamics concerning the validity of
molecular chaos and local equilibrium are violated\cite{Du}. On the other
hand, molecular--dynamics simulations\cite{MD} have the difficulty that
{\em ad hoc }assumptions about microscopic interaction laws have to be
made. An inadequate choice of the interaction parameters can
lead\cite{Luding2} to spurious effects in the simulations.

Here we present a one-dimensional model where colliding
particles interact via an exponential potential. A control parameter
is introduced that allows to perform the limit of a
hard core interaction. On collision particles loose kinetic energy by
the excitation of internal vibrational modes. In a statistical
description we characterize the energy contained in this bath of
internal oscillators  by a bath temperature $T_{bath}$. Our main aim
is to investigate the cooling properties of the system: Starting from
an initial  distribution of center of mass velocities we would like to
understand in detail, how energy is transferred from the translational to the
internal degrees of freedom.

The principal results of this Letter are as follows:
(1) Granular cooling can be understood as energy equipartition among
all (translational and vibrational) modes of the system. (2) At a
single collision the coefficient of restitution $\epsilon$ is a {\em
stochastic}
quantity. We derive an {\em exact} stochastic equation of motion for
the relative coordinates, from which $\epsilon$ can be calculated.
For the special case of "cold" particles ($T_{bath}\!=\!0$) we
recover a classical result of the wave theory of impact. (3) We give
a statistical interpretation of the probability density of $\epsilon$
and propose a new algorithm for simulations of granular particles.

\section*{The model}
We consider a one--dimensional system of $N$
elastic rods with mass density $\rho$ and elastic module $E$.
The elastic state of the $i$th particle   at time $t$ is
described by the longitudinal strain field
$u(s,t),\ -\frac{l_{i}}{2}\leq s\leq
\frac{l_{i}}{2}$, where $s$ is the internal position coordinate and
$l_{i}$ is the length
of the one--dimensional rod. In order to seperate internal and
translational degrees of freedom the strain field contains only fluctuations,
i.e.
$
\int_{-\frac{l_{i}}{2}}^{\frac{l_{i}}{2}}ds\ u_{i}(s,t)\equiv 0
$,
while homogeneous distortions contribute to the center of mass coordinate
$R_{i}$ of the particle. On collision adjacent particles interact via
a potential $V_{hc}(r)$, which depends only on the current end--to--end
distance of the {\em distorted} rods
$r\!=\!R_{i+1,i}+u_{i+1}(-\frac{l_{i+1}}{2},t)
-u_{i}(\frac{l_{i}}{2},t)$, where $R_{i+1,i}\!:=\!R_{i+1}-R_{i}-
\frac{l_{i}+l_{i+1}}{2}$. In the present calculations we investigate
 an exponential interaction potential with  characteristic length scale
$\frac{1}{\alpha}$
\begin{equation}
V_{hc}(r)=B e^{-\alpha r}\quad,
\end{equation}
which includes for large $\alpha$ the limit of a hard core potential.
The constant
$B$ can be chosen arbitrarily, since changing its value merely corresponds
to a rescaling of the relativ coordinates $R_{i+1,i}$. We expand
the strain fields $u_{i}$ in sine and cosine normal modes with
amplitudes
$q_{i}^{(\nu)}(t)$ and normal frequencies $\omega_{i\nu}\!=\!\nu\pi
\frac{c}{l_{i}}$, where $c\!=\!\sqrt{\frac{E}{\rho}}$ is the speed of sound.
If we denote conjugate moments by $P_{i}$ and $p_{i}^{(\nu)}$, respectively,
and treat the elastic energy in harmonic approximation,
the Hamiltonian reads~\cite{Goetzi}:
\begin{eqnarray}
\label{hamilton}
   {\cal H}&=&
{\cal H}_{bath}\left\{p_{i}^{(\nu)},q_{i}^{(\nu)}\right\} + {\cal H}_{trans} +
{\cal H}_{int}\nonumber\\
&=&\sum\limits_{i=1}^{N}\sum\limits_{\nu=1}^{\infty}\left\{
\frac{{p_{i}^{(\nu)}}^{2}}{2 m_{i}}+m_{i}\omega_{i\nu}^{2}
\frac{{q_{i}^{(\nu)}}^{2}}{2}\right\}
+\sum\limits_{i=1}^{N}\frac{P_{i}^{2}}{2 m_{i}}\\
&+&\sum\limits_{i=1}^{N}B\, \exp\!\left\{-\alpha\left(R_{i+1,i}
+\sqrt{2}\sum\limits_{\nu}\left(
q_{i+1}^{(\nu)}-(-1)^{\nu}q_{i}^{(\nu)}\right)\right)\right\}
\quad .\nonumber
\end{eqnarray}
Here we assume periodic boundary conditions, i.e. place the particles
on a ring with perimeter $P$, so that $R_{N+1,N}\!:=\!P+R_{1}-R_{N}$ and
$q_{N+1}^{(\nu)}\!:=\!q_{1}^{(\nu)}$. Because of the nonlinear coupling between
oscillator modes our model has some resemblance with the classical
Fermi--Pasta--Ulam~(FPU) problem~\cite{FPU}.
The important difference is that here the
interaction is "switched" on and off depending on the relative distance of the
translational coordinates: As long as a particle is not involved in a
collision, all its vibrational modes are effectively decoupled.

\section*{Equipartition properties}
We now confine ourselves to two--particle collisions and consider
the case $N\!=\!2$: In the
center of mass frame we introduce the reduced mass $\mu$ and an
effective length scale $l\!=\!2l_{1}l_{2}/(l_{1}+l_{2})$. Taking
$N_{modes}$ normal modes for each particle we numerically
integrate the equations of motion resulting from the
Hamiltonian~(\ref{hamilton}). The following scenario is considered:
We start with an initial relative velocity $\dot
R_{2,1}(0)\!=\!-V_{0}$ and place the particles at maximum distance,
$R_{2,1}(0)\!=\!P/2\gg\frac{1}{\alpha}$. We concentrate on the hard
core limit $\alpha\gg\frac{c}{V_{0}}l$  and set the initial
temperature of the bath of oscillators equal to zero:
$q_{i}^{(\nu)}(0)\!=\!p_{i}^{(\nu)}(0)\!=\!0$. The system now
undergoes an infinite series of collisions with the two particles
colliding alternately at either ends. We observe a rapid decrease of
$E_{tr}\!=\!\frac{\mu}{2}{\dot R_{2,1}}^{\,2}$,
the kinetic energy of the relative coordinate. Fig.~1 displays a decay
of $E_{tr}$ over a range of approximately 100 collisions until a
stationary state is reached with the kinetic energy
fluctuating around an equilibrium value $\mu V_{0}^{2}/2(2
N_{modes}+1)$, that is, of course, essentially zero for large
$N_{modes}$. We interprete this phenomenon as a spreading of the
translational energy among {\em all} degrees of freedom of the
system.

This view is supported by a statistical analysis of the
distribution of vibrational energies:  Denoting modal energies by
$E_{i}^{(\nu)}$ we define the energy of the oscillator bath as
$E_{bath}\!=\!\sum_{i\!=\!1,2}\sum_{\nu}E_{i}^{(\nu)}$. After a
transient time we calculate the relative modal energies
$w_{i}^{(\nu)}\!=\!E_{i}^{(\nu)}/E_{bath}$ after each collision and
take the average. (Due to the continuous interchange of energy
between transational and vibrational modes $E_{bath}$  itself is a
dynamical quantity.)  The resulting energy spectrum (inset of Fig.~1) shows a
homogeneous partition of vibrational energies. This is what we shall
call {\em modal
equipartition regime} in the following. (For particular values of
$\gamma\!=\!l_{1}/l_{2}$, the length ratio of the rods, we found a
periodic structure underlying the spectra. This subtlety does not
affect the main results of this section, however.)
We have also investigated the distribution
function of $w_{i}^{(\nu)}$  ($i$ and $\nu$ fixed)  and always observed an
excellent agreement with the Boltzmann distribution.

An important question is, how fast the modal equipartition regime is
reached. Does the relaxation time (measured by the number of
collisions)
scale with $N_{modes}$? A sensitive probe of equipartition is the
normalized spectral entropy $\eta\!=\!1-h/h_{max}$, where
$h\!=\!-\sum_{i,\nu}w_{i}^{(\nu)}\ln w_{i}^{(\nu)}$ and $h_{max}\!=\!\ln(2
N_{modes})$, which was first introduced in numerical studies on the
FPU problem\cite{Livi}. If equipartition takes place, $\eta$ will
decay from its initial value $\eta\!=\!1$ to its equilibrium value
$\eta_{eq}$ (for Boltzmann distributed $w_{i}^{(\nu)}$ one
finds\cite{Goedde}
$\eta_{eq}\!\approx\!0.423/\ln(2 N_{modes})$). Our plot of $\eta$ for
different numbers of modes (Fig.~2) suggests a relaxation time of
approximately 20 collisions, essentially {\em independent} of
$N_{modes}$. Comparison with Fig.~1 shows that the time scale for modal
equipartition is considerably shorter than the relaxation time of the
granular cooling process.

\section*{Stochastic coefficient of restitution}
 We are now going to
discuss in detail a single two--particle collision in the modal
equipartition regime. Let us denote the precollisional and
post--collisional relative velocity of the particles by $-V$ and $V'$,
respectively. In our model the coefficient of restitution
$\epsilon\!=\!V'/V$ is a stochastic quantity (cf.~Fig.~1) and the
collision process is described by a {\em stochastic} differential
equation. If modal equipartition is valid, we can apply the usual
concepts of statistical physics and consider the pre\-collisional
values of the internal coordinates as independent, Boltzmann
distributed variables (with bath temperature
$T_{bath}\!=\!E_{bath}/2N_{modes}$). Note that this choice of initial
conditions is equivalent to imposing a Markov approximation on the
collisional dynamics, since correlations with preceeding collision
events are neglected.

Obviously, the coefficient of restitution contains the complete
information about the energy transfer between
$E_{tr}\!=\!\frac{\mu}{2}V^{2}$ and $T_{bath}$: Due to conservation of
total energy we have $E'_{tr}\!=\!\epsilon^{2}E_{tr}$ and
$T'_{bath}\!=\!T_{bath}+\frac{1-\epsilon^{2}}{2N_{modes}}E_{tr}$ after
collision.  In order to derive the equation of motion for the
translational coordinate it is convenient to introduce the following
scaled variables: $\tau\!=\!\frac{c}{l}t$ (scaled physical time),\
$z(\tau)\!=\!\alpha R_{2,1}$, $z_{0}\!=\!z(\tau_{0})$ (initial
distance), and $\kappa\!=\!\alpha l\frac{V}{c}$ ($\!=\!-\frac{d z}{d
  \tau}(\tau_{0})$, scaled initial velocity). Again we are mainly
interested in the hard core limit so that now $\kappa\!\gg\!1$ is our
large control parameter. Furthermore we set the free constant
$B\!=\!\mu\frac{c^{2}}{(\alpha l)^{2}}$. The internal coordinates in
Hamilton's equations of motion can be integrated out\cite{Goetzi}
using Green's function method and we arrive at the following equation
for $W(\tau)\!=\!\frac{d z}{d \tau}+\kappa$, the {\em velocity
  increase} throughout collision:
\begin{eqnarray} \label{SDE}
\frac{d W}{d\tau}
&=&\exp\left\{\kappa\tau-z_{0}-W(\tau) -\sum\limits_{i=1,2}
\sum\limits_{n=1}^{\infty}W(\tau-n\Gamma_{i})
+Q_{1}(\tau)+Q_{2}(\tau)
\right\}\quad,
\end{eqnarray}
with $W(\tau)\!:=\!0\mbox{\quad for
}\tau\le \tau_{0}$. The coefficient of restitution is then obtained
from the asymptotics  of $W$:
\begin{equation}
   \epsilon=\lim_{\tau\to\infty} \frac{W(\tau)}{\kappa} - 1\quad.
\end{equation}
In Eq.~(\ref{SDE}) $\Gamma_{1}\!=\!1+\gamma$ and
$\Gamma_{2}\!=\!1+1/\gamma$ are the periods of the lowest normal
modes in scaled time and $Q_{1},Q_{2}$ are Gaussian stochastic processes
resulting from the initial conditions. They take on the form of
stochastic Fourier series
$Q_{i}(\tau)\!=\!\sum_{n=1}^{\infty}\frac{a_{in}}{n}\sin(n
2\pi\tau/\Gamma_{i}) + \sum_{n=1}^{\infty}\frac{b_{in}}{n}\cos(n
2\pi\tau/\Gamma_{i})$ with independent and identically distributed
Gaussian coefficients $a_{in},b_{in}$ .
As a consequence, we have
\begin{eqnarray}
\label{CF}
   <Q_{i}(\tau)>&=&0\\
<Q_{i}(\tau)Q_{j}(\tau')>&=&\delta_{ij}\,\kappa^{2}\frac{T_{bath}}{E_{tr}}
\frac{\Gamma_{i}}{4}\left[\left(\frac{\tau-\tau'}{\Gamma_{i}}-\frac{1}{2}
\right)^{2}
-\frac{1}{12}\right]\nonumber
\end{eqnarray}
for $0\!\le\!\tau-\tau'\!\le\!\Gamma_{i}$ and periodic continuation everywhere
else. The periodicity of the correlation function and the memory terms
in Eq.~(\ref{SDE}) reflect the fact, that we consider a system of {\em ideal}
oscillators that never forgets its collision history. Note that
typical fluctuations of the noise processes are of the same order as
$W(\tau)$, which makes an analytical treatment
even more difficult.

However, it is possible to solve the deterministic equation
($T_{bath}\!=\!0$): In the large $\kappa$ regime the coefficient of
restitution is essentially given by the length ratio $\gamma$:\
$\epsilon\!=\!\min(\gamma,1/\gamma)\!+\!\ln(\kappa|\Gamma_{2}\!-\!
\Gamma_{1}|)/\kappa$, and the actual collision takes place in the time
intervall
$\lbrack\frac{z_{0}}{\kappa},\frac{z_{0}}{\kappa}\!+\!\min(\Gamma_{1},
\Gamma_{2})\rbrack$, i.e. the duration of the collision is equal to
the wave propagation time of the shorter rod.  These results are in
agreement with the wave theory of impact~\cite{Gold}, that analyzes
the collision problem in terms of stress waves propagating through the
rods. To our knowledge, a derivation in the framework of classical
mechanics (with correction terms for finite $\kappa$) has not been
given before.

For $T_{bath}\!\ne\!0$ we had to integrate Eq.(\ref{SDE}) numerically
and average over many realizations of the noise processes. We have
focussed on the statistics of $\epsilon^{2}$, since it is the
relevant quantity for energy transfer. In the hard core regime
($\kappa$ larger than $\approx\!30$)  the probability density of
$\epsilon^{2}$, $p(\epsilon^{2})$, does not significantly depend on
$\kappa$, but it is still a function of $E_{tr}/T_{bath}$ and
$\gamma$. Recall that Eq.~(\ref{SDE}) describes a {\em single}
collision, given the {\em current} values of $E_{tr}$  and $T_{bath}$.
After collision the energy ratio has changed to $E'_{tr}/T'_{bath}$
(see above) and in turn there is a different distribution of
$\epsilon^{2}$ at the next collision. Fig.~3 displays plots of the
distribution function for  different values of the parameters.

The important point is that $p(\epsilon^{2})$
has to be interpreted as a {\em transition density}: In the context
of the Markov approximation leading to Eqs.(\ref{SDE}),(\ref{CF}),
the time evolution of $E_{tr}$ (cf. Fig.~1) is simply a Markovian  jump
process in discrete "time" (i.e. cumulative number of collisions).
If the system is in a state with energy $E_{tr}$, the probability
density for a transition to energy $E'_{tr}$ is determined by
\begin{equation}
 p(E_{tr}\rightarrow E'_{tr}) =  \frac{1}{E_{tr}}
p\!\left(\epsilon^{2}=\frac{E'_{tr}}{E_{tr}}\right)\quad.
\end{equation}
As expected, the transition density resulting from
Eq.(\ref{SDE}) can be shown numerically to yield a stationary energy density
$p^{stat}(E_{tr})\!\propto\!\exp(-E_{tr}/T_{bath})$ . In view of the
many--particle problem (see below) it seems reasonable to replace
the exact distribution
of $\epsilon^{2}$ by a simpler one, that obeys detailed balance and
leads to the same stationary density $p^{stat}(E_{tr})$. A possible
choice, for example, is  the analog of
Glauber dynamics\cite{Glauber} for Ising spin systems:
\begin{equation}
p^{(G)}(\epsilon^{2})=\frac{E_{tr}}{T_{bath}}
\exp\!\left(-\epsilon^{2}\frac{E_{tr}}{T_{bath}}\right)\quad.
\end{equation}

\section*{Many--particle problem}
 For the many--particle system
translational energy and bath temperature become functions of time
{\em and} position.  From the preceeding sections it is clear that
each two--particle collision tends to establish a state where {\em
  locally} $E_{tr}\!\approx\!T_{bath}$. As in the case $N=2$, we
assume modal equipartition for each particle and treat collisions in
the Markov approximation. Furthermore, we expect that the cooling
properties of the system will not sensitively depend on the detailed
form of the probability density $p(\epsilon^{2})$.

For computer simulations we therefore suggest an algorithm that is a
combination of ED and Monte Carlo methods: Each of the $N$ particles
is assigned an individual bath temperature $T^{(i)}_{bath}$ that
characterizes its vibrational state. For each collision the following
steps have to be performed: (1) Calculate the next collision event by
the ED procedure. (2) For the colliding pair determine $E_{tr}$ and
an averaged local bath temperature $T_{bath}$. (3) With these
parameters choose $\epsilon$ randomly
from a simple distribution, e.g.\ $p^{(G)}(\epsilon^{2})$. (4) Update
the relative velocities and the post--collisional values of
$T^{(i)}_{bath}$; go back to (1).

Another possible extension of our study would be a generalization of
the inelastic Boltzmann equation \cite{McNamara,Bernu} to a
stochastic coefficient of restitution.  Work along these lines is in
progress.

\section*{Acknowledgments}
This work has strongly benefitted from the
  cooperation and many helpfull discussions with Annette Zippelius and
  Reiner Kree. Support by the Sonderforschungsbereich~345 is
  gratefully acknowledged.

\newpage
\section*{Figure captions}
\begin{quote}
{\bf Figure~1:} Time evolution of the relative kinetic energy for a
  two--particle system with length ratio
  \protect$\gamma\!=\!l_{1}/l_{2}\!=\!0.6173$, \protect$\alpha V_{0}/c
  l\!=\!4000$, and \protect$N_{modes}\!=\!75$.  The dashed line is the
  equilibrium value resulting from energy equipartition.\protect\\ The
  inset displays a cutout from the corresponding spectrum of modal
  energies: After a transient time of 500
  collisions the relative energies \protect$w_{i}^{(\nu)}$ have been
  averaged over another 4000 collisions. Grey and black bars
  correspond to particle 1 and 2, respectively.
\end{quote}
\bigskip
\begin{quote}
{\bf Figure 2:} Time behaviour of the spectral entropy \protect$\eta$ for
different numbers of modes. Again, \protect$\gamma\!=\!0.6173$ and
\protect$\alpha V_{0}/c l\!=\!4000$.
\end{quote}
\bigskip
\begin{quote}
{\bf Figure~3:} The probability density \protect$p(\epsilon^{2})$ (dotted
line) and its integral, the distribution function (full line), for
different energy and length ratios: a) \protect$E_{tr}/T_{bath}\!=\!10$,
$\gamma\!=\!0.6173$,\ b)\protect~$E_{tr}/T_{bath}\!=\!1$,
$\gamma\!=\!0.6173$,\ and c) \protect$E_{tr}/T_{bath}\!=\!1$,
$\gamma\!=\!0.2173$.
Data were obtained from the numerical integration of 8000 realizations of
Eq.~(\protect\ref{SDE}) in the hard core regime (\protect$\kappa\!=\!100$).
\end{quote}

\end{document}